# Frustrated ferromagnetic transition in AB-stacked honeycomb bilayer

S. Y. Wang[1#], Y. Wang[1#], S. H. Yan[2#], C. Wang[2#], B. K. Xiang[1], K. Y. Liang[1], Q. S. He[1], K. Watanabe[3], T. Taniguchi[4], S. J. Tian[2], H. C. Lei[2], W. Ji[2], Y. Qi[1], Y. H. Wang[1,5*]

1. State Key Laboratory of Surface Physics and Department of Physics, Fudan University, Shanghai 200438, China
2. Department of Physics and Beijing Key Laboratory of Opto-electronic Functional Materials & Micro-nano Devices, Renmin University of China, Beijing 100872, China
3. Research Center for Functional Materials, National Institute for Materials Science, 1-1 Namiki, Tsukuba 305-0044, Japan
4. International Center for Materials Nanoarchitectonics, National Institute for Materials Science, 1-1 Namiki, Tsukuba 305-0044, Japan
5. Shanghai Research Center for Quantum Sciences, Shanghai 201315, China

# These authors contributed equally.

* To whom correspondence and requests for materials should be addressed. Email: wangyhv@fudan.edu.cn

**Abstract:** In two-dimensional (2D) ferromagnets, anisotropy is essential for the magnetic ordering as dictated by the Mermin-Wagner theorem. But when competing anisotropies are present, the phase transition becomes nontrivial. Here, utilizing highly sensitive susceptometry of scanning superconducting quantum interference device microscopy, we probe the spin correlations of ABC-stacked $CrBr_3$ under zero magnetic field. We identify a plateau feature in susceptibility above the critical temperature ($T_C$) in thick samples. It signifies a crossover regime induced by the competition between easy-plane intralayer exchange anisotropy versus uniaxial interlayer anisotropy. The evolution of the critical behavior from the bulk to 2D shows that the competition between the anisotropies is magnified in the reduced dimension. It leads to a strongly frustrated ferromagnetic transition in the bilayer with fluctuation on the order of $T_C$, which is distinct from both the

**monolayer and the bulk. Our observation potentially offers a 2D localized spin system on honeycomb lattice to explore magnetic frustration.**

The honeycomb lattice requires exotic nearest-neighbor interactions to generate frustration. This is because its bipartite unit cell avoids antiferromagnetic frustration by an alternate up-down arrangement of spins [1]. Kitaev proposed a bond-dependent nearest-neighbor ferromagnetic interaction to generate frustration on the honeycomb lattice [2] [3]. Realization of the Kitaev model in realistic planar materials is hotly sought-after [4] but highly challenging [1,5]. The recent advent of van der Waals magnetic materials [6-10] has provided a new playground for studying spin correlation in true 2D systems. Their magnetic ground states are highly tunable by magnetic field [11], pressure [12], or stacking order [13-21]. These knobs are changing the dominant magnetic anisotropy within and between the layers essential for the ordering in 2D. But when two anisotropies from different sublattices with similar magnitude and opposite signs compete, they provide suitable ingredients for magnetic frustration on a honeycomb lattice. However, the 2D magnetic ordering and phase transition under such competing anisotropies has rarely been explored.

As van der Waals semiconductors with large bandgaps, the chromium trihalide family is representative as a localized spin system on a honeycomb lattice. Each van der Waals layer of $CrBr_3$ is composed of one layer of $Cr^{3+}$ ions in a honeycomb network sandwiched between two layers of $Br^-$ ions (Fig. 1a) [22]. The super-exchange interaction through $Br^-$ between the intralayer nearest-neighbor spins on $Cr^{3+}$ favors parallel correlation ($J_1 > 0$), which makes the monolayer ferromagnetic. The next nearest-neighbor interaction, the interlayer super-superexchange ($J_2$), determines the magnetic ground state of multi-layered structures depending on the stacking [14,23-30]. CrBr$_3$ is special in its family that it maintains the rhombohedral structure from bilayer to the bulk and from low temperature to room temperature [7]. For such a stacking, the interlayer exchange is also ferromagnetic ($J_2 > 0$). Magnetic anisotropy is essential for the ordering of the localized spins in the 2D limit according to the Mermin-Wagner theorem. Thus, we consider a general anisotropic Heisenberg model [31] represented by the Hamiltonian:

$$H = -\frac{1}{2}\sum_{\langle ii'\rangle}^{a=1,2} J_a S_i \cdot S_{i'} - \frac{1}{2}\sum_{\langle ii'\rangle}^{a=1,2} \Lambda_a S_i^z S_{i'}^z - D\sum_i (S_i^z)^2$$

where $a = 1,2$ denotes taking the intralayer or the interlayer parameter (Fig. 1), respectively, depending on the two neighboring spins ($S_i$ and $S_{i'}$) involved in the summation; $\Lambda_1$ and $\Lambda_2$ are the exchange anisotropies, $D$ is the single-ion anisotropy; $z$ is the easy-axis for positive values of the anisotropies.

These intra- and inter-layer anisotropies are the fundamental factors for frustration in a ferromagnetic Heisenberg model on a honeycomb lattice. When the anisotropies have the same sign or one of them is much bigger than the others, as is the case of (bulk) CrI$_3$, this Hamiltonian is similar to an Ising model. But when $\Lambda_1$ and $\Lambda_2$ have the opposite sign and similar magnitude, their competition may cause frustration on the honeycomb lattice. Particularly for AB-stacked CrBr$_3$ bilayer (Fig. 1b), interlayer uniaxial anisotropy ($\Lambda_2 > 0$) on the A sites is frustrated by easy-plane anisotropy from nearest-neighbor bonds ($\Lambda_1 < 0$). And $\Lambda_1 + \Lambda_2$ is only about 7 μeV/Cr according to our first-principle calculation (SOM). It is an order of magnitude smaller than that of CrI$_3$, where all the anisotropies have the same sign favoring the easy-axis [26].

Since CrBr$_3$ bilayer is known to order ferromagnetically at low temperatures, experimental investigation of the magnetic frustration in this system hinges on probing spin correlation in the critical regime. According to the fluctuation-dissipation theorem, spin-spin correlation is directly reflected in magnetic susceptibility [32]. Volumetric techniques for measuring magnetic susceptibility and studying frustration in bulk materials such as neutron scattering [33] is not applicable to 2D materials. On the other hand, magnetic microscopy relying on optical means, though sensitive to magnetization, are not capable of measuring susceptibility. Furthermore, external magnetic field applied in these techniques to enhance magnetization and obtain hysteresis loop may alter the magnetic ground state and the critical behavior at the phase transition [13,16,34-36]. Scanning SQUID (sSQUID) microscopy is a sensitive and direct imaging technique for detecting weak magnetic signal without the need of external field [37-39]. Nano-fabricated SQUID equipped with a field coil is especially suitable to perform local susceptometry [40-42].

We first identify the unique signature of competing intra- and inter-layer magnetic interactions in CrBr$_3$ bulk crystals. The sSQUID magnetometry $\Phi(T)$ of a thick flake under zero field shows magnetic domains with opposite out-of-plane magnetization at 12 K (Fig. 2a). The contrast is qualitatively the same till 27 K, above which it suddenly vanishes. The variation in the specific domain pattern from different cooling cycles (Fig. 2a, $T$ = 27 K) suggests the domains are not pinned by disorder, which is consistent with little magnetic hysteresis in bulk magnetometry (Fig. S2) [23,43,44]. Similar domain patterns below 27 K also suggests that the absence of magnetization at higher temperatures is not due to a much-reduced domain size beyond our resolution.

The susceptometry reveals spin correlations at the critical regime not accessible by the magnetometry. The alternate-current susceptibility $\chi'$ images (Fig. 2b) show uniform and similar paramagnetic signal at 12 K and 20 K uncorrelated to magnetization. There occurs a strong response at 27 K and then quickly reduces at higher temperatures. Fixing on one location for finer temperature variation, we find two distinctive features in $\chi'(T)$: a spike at 28.2 K and a plateau regime starting around 32 K (Fig. 2c). Such $\chi'(T)$ is in stark contrast to that of CrI$_3$ (Fig. S7), which does not have a plateau regime and is more similar to a conventional ferromagnet [11]. The spike at 28.2 K in CrBr$_3$ coincides with the onset temperature of its out-of-plane magnetization (Fig. 2c, blue), signifying a ferromagnetic phase transition at this $T_C$. The critical exponent at $T_C$ as determined from fitting of the $\Phi(T)$ using Curie-Weiss function was $\beta = 0.094$, again not matching any of the spin models without considering interlayer interaction. But the entire $\chi'(T)$ can be well fit by an empirical piecewise function composed of three Brillouin functions (Fig. 2c, orange), giving the susceptibility peak temperature $T_P = 28.2$ K and plateau temperature $T_X = 33.6$ K. The susceptibility feature at $T_P$ is suppressed by a field of 150 Oe, which agrees with the small energy scale of interlayer coupling [31,45,46].

Our Monte-Carlo simulation finds that suppression of magnetic ordering occurs in the parameter space where $\Lambda_1$ and $\Lambda_2$ competes. For easy-plane intralayer anisotropy ($\Lambda_1 < 0$) and uniaxial interlayer anisotropy ($\Lambda_2 > 0$), there is no spike in susceptibility when the two are roughly matched $\Lambda_1 + \Lambda_2 \sim 0$ and $D$ is small (Fig. 2d). There is no magnetization at finite temperature (Fig. 2e) and specific heat has no spike either (SOM), indicating the absence of out-of-plane ferromagnetic transition. Instead, there is a plateau in $\chi'(T)$ around 35 K. At larger $D$, we find a spike in $\chi'(T)$ below the plateau simultaneously with onset of magnetization, indicating the return of ferromagnetic order. $T_C$ increases with increasing $D$ so that the plateau width shrinks.

At $D = 0.446$, the two characteristic temperatures from the simulation match the $T_C$ and $T_X$ from the sSQUID measurements (Fig. 2c). At even larger $D > 0.46$, the susceptibility plateau disappears and the critical behavior follows a typical Ising transition again. The small range of $D$ with the plateau behavior is a manifestation of magnetic frustration above the phase transition.

Having identified the competing nature between the magnetic anisotropies in the bulk, we next examine their impact in the 2D limit. We compare side-by-side a bilayer with a monolayer, which does not have any interlayer interaction (Fig. 3). We have obtained 87 sSQUID images as a function of rising temperature from 13 K to 40 K after zero field cooling. The representative magnetometry images show single domain magnetization for both the monolayer and the bilayer with stronger signal for the latter (Fig. 3a). Their magnetizations become weaker as the temperature rises from 13 K to 34 K but the temperature at which the magnetization diminishes are different: 24 K for the monolayer and 28 K for the bilayer. The $T_c$ of the bilayer is 8 K higher than that of the monolayer and similar to those of the bulk (Fig. 2) and multilayers (SOM). This is consistent with the previous observation by magnetic circular dichroism [17] and Hall magnetometry [18]. But we note that $T_c$'s are slightly lower in our case because we do not apply any magnetic field to measure magnetization [13,16,17].

The susceptometry exhibits distinctive spin fluctuation in the bilayer from the monolayer not discernable in magnetometry. The susceptometry images at the lowest measuring temperature of 13 K does not show any contrast on the monolayer while the bilayer shows clear $\chi'$ (Fig. 3b). In contrast to the magnetization, the $\chi'$ becomes stronger as temperature increases to 15 K and 20 K when the monolayer is visible. The monolayer's $\chi'$ disappears again at 24 K while the bilayer's gets even stronger. $\chi'$ for the bilayer finally gets weaker at 27 K, the temperature at which its magnetization already diminishes, and completely disappears at 34 K.

The temperature dependence of monolayer magnetization and susceptibility shows no sign of a crossover (Figs. 3c). Averaging over the areas of the monolayer, we find $\Phi(T)$ can be barely fit with a Curie-Weiss function $M \propto (T_c - T)^\beta$ with the critical exponent $\beta = 0.46$ (Fig. 3c, dashed purple), which is consistent with the $\beta$ obtained in the Hall magnetometry [18]. It is much higher than the theoretical 2D Ising case ($\beta = \frac{1}{8}$) but strangely close to the mean-field value [8]. The non-Ising $\beta$ suggests the intralayer anisotropy does not have a conventional easy-axis, which is consistent with our observation in the bulk. There is a singular susceptibility peak centered around $T_P = 20$ K, which coincides with the onset of magnetization at $T_c$ (Fig. 3c) [8]. Unlike in thicker samples, there is no visible plateau in $\chi'(T)$. Again, $\chi'(T)$ is poorly fitted by a power law but well-fitted by a piece-wise function composed of two Brillouin functions, yielding $T_C = T_P = 20.2$ K (Fig. 3c, orange). Monolayer's lack of the susceptibility plateau above $T_C$ directly reflects the absence of competing interlayer interaction.

The situation is quite different in the bilayer. Its $\chi'(T)$ shows a very broad peak that rises almost linearly on the high temperature side except for a small hump around 27 K (Fig. 3d). By fitting it with three piecewise Brillouin functions, we obtain $T_P = 22.8$ K and $T_X = 27.7$ K. $\Phi(T)$ increases slowly below $T_c = 28$ K until tailing off around 15 K. A much lower $T_P$ than $T_C$ suggests the ferromagnetic transition in the bilayer is highly nontrivial. It is neither of the Ising type as seen in the bulk (Fig. 2c), nor close to the 'mean-field' kind in the monolayer (Fig. 3c). The broad $\chi'(T)$ peak suggests spin fluctuation extends well beyond the critical regime of the bilayer.

In order to understand the critical behavior of the bilayer, we vary the layer number to examine how susceptibility evolves from a higher lattice dimension. As the thickness reduces, the overall line-shape of $\chi'(T)$ gradually changes from sharp to rounded (Fig. 4a), except for the monolayer. The peak in $\chi'(T)$ centered at $T_P$ continuously broadens with reducing $N$ (Fig. 4b). The transition width $\Delta T$, which is obtained from the Brillouin fit, follows a power law with $N^{-1}$. It reaches a maximum of $\Delta T = 23$ K at the bilayer before turning down sharply at the monolayer (Fig. 4c). For comparison, the $\chi'(T)$ in the logarithm scale shows a constant slope above $T_X$ independent of $N$ (Fig. 4d). This quasi-2D paramagnetic behavior is consistent with our anisotropic Heisenberg model where the intralayer interaction dominates around $T_X$. Taken together with the spin correlation in the bulk (SOM), such contrast suggests that the interlayer interaction is responsible for the increasing spin fluctuation with reducing $N$ around $T_P$.

Although the bilayer and the bulk both have a ferromagnetic ground state, their ferromagnetic transitions belong to different universality classes. From the bulk limit to $N = 150$, both $T_X$ and $T_P$ are independent of $N$ and $T_P = T_C$; and below $N = 150$, both $T_X$ and $T_P$ reduce, making $T_P < T_C$ (Fig. 4e). The reduction of these two temperatures occurs with enhanced fluctuations around $T_P$ so that the rising of the susceptibility peak is always around $T_C$. Such $N$ dependence of the characteristic temperatures and fluctuation result in the merge of the plateau and the spike. The collapse of the crossover regime for $N < 30$ suggests that the phase transition transforms to a 2D universality class. This is notably different from the dimensional crossover in $Fe_3GeTe_2$ which is represented by the drastic reduction of $T_C$ [10,47]. The fluctuation around the transition cannot be simply explained by the reduced dimensionality because the fluctuation in the monolayer is much smaller (Fig. 3c). Other distinctions of the monolayer are that its out-of-plane spin correlation peaks at $T_C$ and its $T_C$ is much lower than the multilayers. These facts suggest that the phase transition in the monolayer belongs to yet another 2D universality class.

Conversely, the fluctuation in the multilayers deep into the 2D ferromagnetic order provides evidence for the frustration induced by the competing interlayer anisotropy. Recall that in the 3D limit the total easy-axis anisotropy ($\Lambda_2 + D$) is slightly stronger than the easy-plane anisotropy $\Lambda_1$ such that an out-of-plane ferromagnetic order wins out at $T_C$ after competing throughout the crossover. As the number of layers reduces, fluctuation on the interlayer exchange is enhanced, effectively weakening $\Lambda_2$, and therefore the competition between the two anisotropies becomes stronger. Overall, the reduced dimensionality magnified the competition satisfied by $\Lambda_1 + \Lambda_2 + D \sim 0$. The frustration culminates in the bilayer where $\Delta T$ is on the same order as $T_c$, manifesting a heavily frustrated ferromagnetic transition.

The nature of the spin fluctuation in the bilayer remains to be investigated. The tunability of 2D van der Waals materials offers many knobs to experimentally access the parameter space of $\Lambda_1$, $\Lambda_2$ and $D$. For example, slightly increasing $D$ may recover an Ising transition while reducing it may realize a strongly frustrated spin system with a large frustration parameter [3]. The reduced energy scales in frustrated spin systems [1,4,5] is sensitive to a small external field. The direct flux sensitivity and spatial resolution of sSQUID under zero applied field can be useful for detecting collective spin excitations in 2D materials.

In conclusion, we have identified competing intra-layer and inter-layer anisotropies at the critical regime in CrBr$_3$ by employing sSQUID susceptometry. There is a 2D-3D crossover regime of constant short-range magnetic correlation above $T_C$ in thick samples. This regime shrinks while spin fluctuation increases as a function of reducing layer thickness, indicating a transition to 2D universality class. In the bilayer, the spin fluctuation covers a broad temperature range on the same order as $T_C$, suggesting a frustrated ferromagnetic transition. Our methodology may be generally applied to the study of magnetic frustration in van der Waals materials.

## Acknowledgement

YHW would like to acknowledge partial support by the Ministry of Science and Technology of China (Grant No. 2017YFA0303000), National Natural Science Foundation of China (Grant No. 11827805 and 12150003), National Key R&D Program of China (Grant No. 2021YFA1400100) and Shanghai Municipal Science and Technology Major Project (Grant No. 2019SHZDZX01). HCL would like to acknowledge support by National Key R&D Program of China (Grant No. 2018YFE0202600), Beijing Natural Science Foundation (Grant No. Z200005). K.W. and T.T. acknowledge support from JSPS KAKENHI (Grant Numbers 19H05790, 20H00354 and 21H05233) and A3 Foresight by JSPS. YQ acknowledges financial support from NNSFC Grant No. 11874115. W.J gratefully acknowledges financial support from the Ministry of Science and Technology (MOST) of China (Grant No. 2018YFE0202700), the National Natural Science Foundation of China (Grants No. 11974422), the Strategic Priority Research Program of Chinese Academy of Sciences (Grant No. XDB30000000), the Fundamental Research Funds for the Central Universities, China, and the Research Funds of Renmin University of China (Grants No. 22XNKJ30). C.W was supported by the National Natural Science Foundation of China (Grants No. 12104504) and the China Postdoctoral Science Foundation (2021M693479). Calculations were performed at the Physics Lab of High-Performance Computing of Renmin University of China, Shanghai Supercomputer Center. All the authors are grateful for the experimental assistance by D. L. Feng, T. Zhang, S. Y. Li and L. F. Yin and for the stimulating discussions with S. W. Wu, C. L. Gao, Y. Z. Wu, J. Zhao, Y. Wan and P. Z. Tang.

## Data availability

The data that support the findings of this work are available from the corresponding author upon reasonable request.

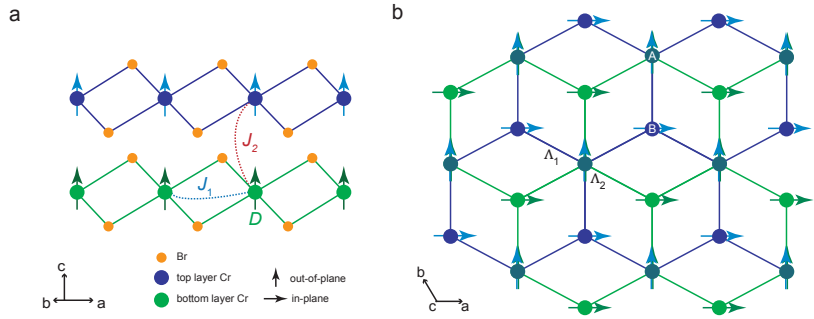

**Fig. 1 Crystal structure of AB-stacked CrBr$_3$ bilayer and intra- and inter-layer magnetic interactions. a**, Side-view of a bilayer structure. The spins are localized on the Cr sites and interact with each other via Br. $J_1$ and $J_2$ are the exchange interactions between the intralayer and interlayer nearest-neighbor spins, respectively; $D$ is single-ion anisotropy. Both $J_1$ and $J_2$ are ferromagnetic in CrBr$_3$ with the illustrated stacking. **b**, Top-view of the AB-stacked honeycomb. (Br's are omitted for clarity.) $\Lambda_1$ and $\Lambda_2$ are the intra- and inter-layer exchange anisotropies, respectively. CrBr$_3$ favors an easy-plane intralayer anisotropy ($\Lambda_1 < 0$), and a uniaxial interlayer anisotropy ($\Lambda_2 > 0$), which is effective only on 'A' sites. When $\Lambda_1 + \Lambda_2 \sim 0$ with a small $D$, the spins are frustrated.

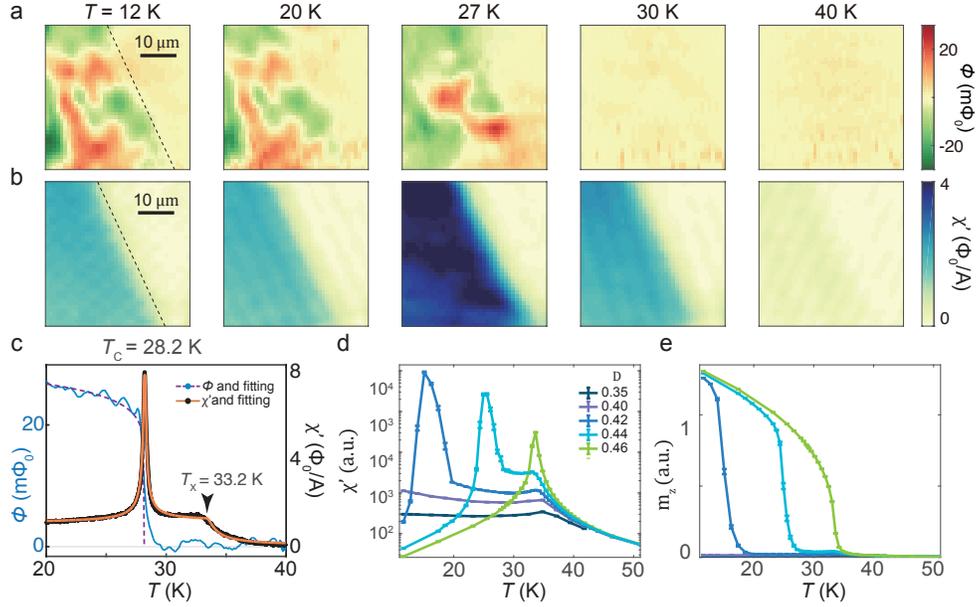

**Fig. 2 Susceptibility plateau above the ferromagnetic phase transition in a thick CrBr$_3$ crystal. a**, Magnetometry ($\Phi$) images at various temperatures $T$ after zero-field cooling. The image at $T = 27$ K is obtained in a different thermal sequence, leading to a different domain pattern. The dashed line marks the boundary between the 150-nm-thick sample (left) and substrate (right). **b**, Alternate-current susceptometry ($\chi'$) images obtained simultaneously with magnetometry. **c**, $\Phi(T)$ and $\chi'(T)$ measured by sSQUID at a fix location on the sample. Magnetization critical temperature ($T_C$) coincides with the spike in susceptibility, $T_P = T_C = 28.2$ K. The dashed-line is a fitting of $\Phi(T)$ using the Curie-Weiss function $(T_c - T)^\beta$, yielding $\beta = 0.094$. The orange line is a fit of $\chi'(T)$ using piece-wise Brillouin functions. The plateau corresponds to another characteristic temperature $T_X = 33.2$ K. **d** and **e**, Monte-Carlo simulation of the ferromagnetic susceptibility and out-of-plane average magnetization, respectively, at different $D$. The other parameters are: $J_1 = 2.98$, $J_2 = 0.125$, $\Lambda_1 = -0.298$, $\Lambda_2 = 0.085$ (all in meV). The simulation which reproduces the experimental $T_X$ and $T_P$ occurs at $D = 0.447$.

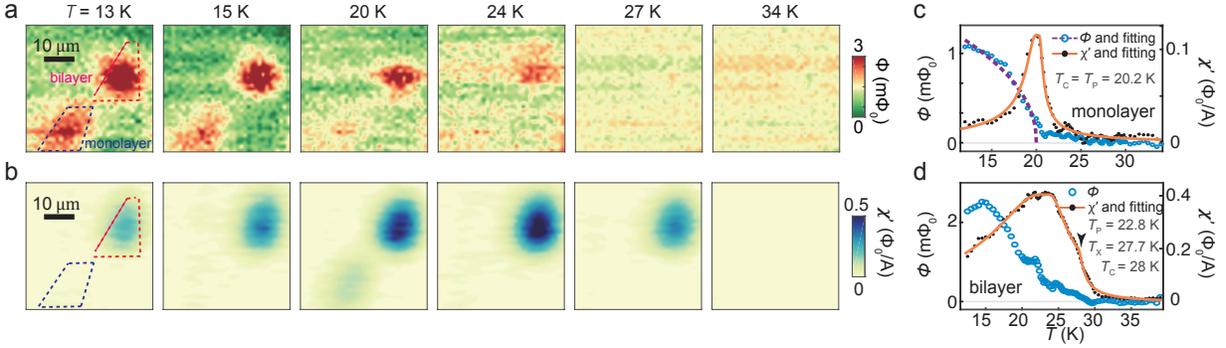

**Fig. 3 sSQUID magnetometry and susceptometry images of monolayer and bilayer CrBr$_3$ across the critical regime. a,** Magnetometry images at various $T$. The red dashed-line and the purple dashed-line mark the bilayer and the monolayer determined from the optical image, respectively. **b,** Corresponding susceptometry images obtained simultaneously as the magnetometry images. **c** and **d,** $\Phi(T)$ (left axis) and $\chi'(T)$ (right axis) extracted from 87 images at different temperatures by averaging over the monolayer and bilayer regions, respectively. The purple dashed-line in **c** is a Curie-Weiss fit of $\Phi(T)$. The orange lines are fits of $\chi'(T)$ using piece-wise Brillouin functions. Note that $T_P$ coincides with $T_C$ in the monolayer but not in the bilayer.

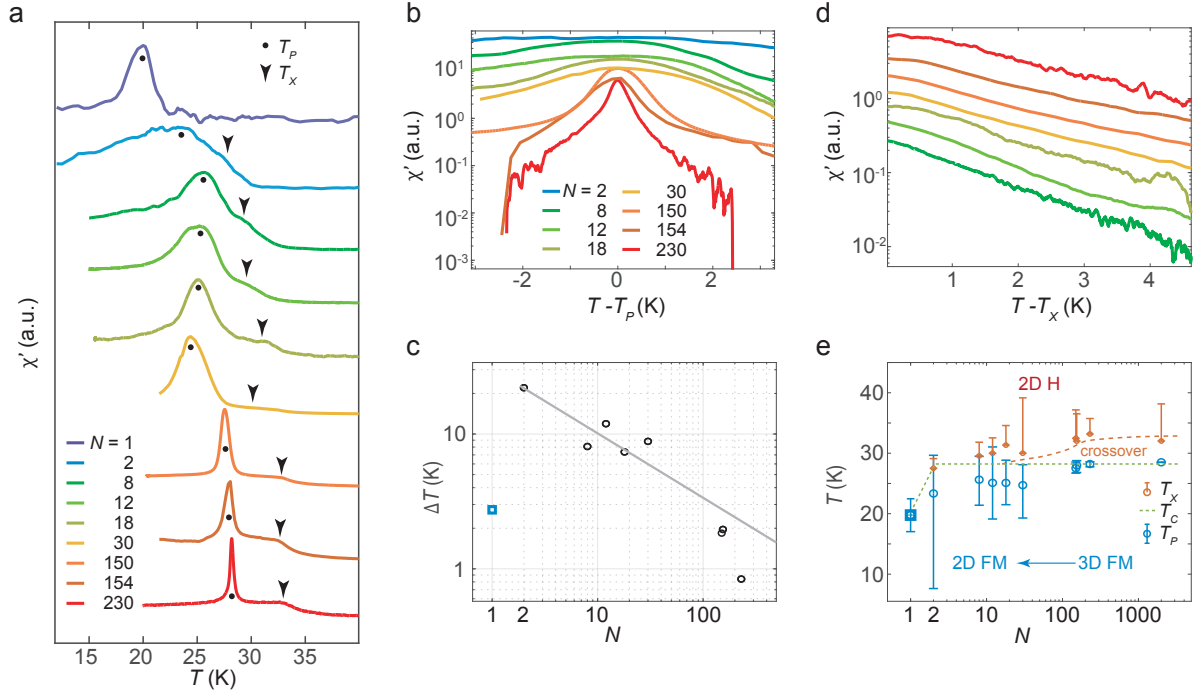

**Fig. 4 Evolution of the frustrated ferromagnetic transition of CrBr$_3$ from the 3D limit to the bilayer. a,** $\chi'(T)$ of samples with different layer number $N$ at zero field (rescaled vertically). **b,** Peaks in $\chi'(T - T_P)$ in logarithm scale after subtracting the plateau. **c,** Width of the peaks as a function of $N$. Gray line: power-law fit for $N \geq 2$. **d,** $\chi'(T - T_X)$ in logarithm scale showing constant slope above $T_X$. **e,** Phase diagram in terms of $N$. $T_P$ and $T_X$ are determined from Brillouin-function-fitting of $\chi'(T)$; $T_C$ are from the onset in $\Phi(T)$. The error bars on $T_P$ and $T_X$ are the transition width from the fittings, representing the range of spin fluctuation. '2D H' is a paramagnetic phase without interlayer correlation. The orange dashed-line is a guide-to-the-eye. Frustration induced by the competing intra- and inter-layer anisotropies is magnified in the 2D transition, culminating at the bilayer with transition width approaching $T_C$.